\documentclass[aps,prl,twocolumn]{revtex4}

\usepackage{amsmath}
\usepackage{amssymb}
\usepackage{stmaryrd}
\usepackage{bm}
\usepackage{graphicx}

\usepackage{epsfig}

\def\ov#1{\overline{#1}}

\def\vb#1{\mbox{\boldmath$#1$}}
\def\pd#1#2{\frac{\partial #1}{\partial #2}}
\def\fd#1#2{\frac{\delta #1}{\delta #2}}
\def\wh#1{\widehat{#1}}
\def\bdot{\,\vb{\cdot}\,}
\def\btimes{\,\vb{\times}\,}

\def\bhat{\wh{{\sf b}}}
\def\cal#1{\mathcal{#1}}

\def\bhat{\wh{{\sf b}}}

\newcommand{\bc}{\begin{center}}
\newcommand{\ec}{\end{center}}
\newcommand{\bt}{\begin{tabbing}}
\newcommand{\et}{\end{tabbing}}
\newcommand{\be}{\begin{eqnarray*}}
\newcommand{\ee}{\end{eqnarray*}}
\newcommand{\bs}{\begin{slide}}
\newcommand{\es}{\end{slide}}

\begin{document}

\title{Gauge-free electromagnetic gyrokinetic theory}

\author{J.~W.~Burby$^{1,3}$ and A.~J.~Brizard$^{2}$}
\affiliation{$^{1}$Courant Institute of Mathematical Sciences, New York, NY 10012, USA \\ $^{2}$Department of Physics, Saint Michael's College, Colchester, VT 05439, USA\\ $^3$Los Alamos National Laboratory, Los Alamos, NM 87545, USA}

\begin{abstract}
A new gauge-free electromagnetic gyrokinetic theory is developed, in which the gyrocenter equations of motion and the gyrocenter phase-space transformation are expressed in terms of the perturbed electromagnetic fields, instead of the usual perturbed potentials. Gyrocenter polarization and magnetization are derived explicitly from the gyrocenter Hamiltonian, up to first order in the gyrocenter perturbation expansion. Expressions for the sources in Maxwell's equations are derived in a form that is suitable for simulation studies, as well as kinetic-gyrokinetic hybrid modeling. 
\end{abstract}

\date{\today}

%\pacs{}% insert suggested PACS numbers in braces on next line

\maketitle

Nonlinear gyrokinetic theory provides a self-consistent description of low-frequency kinetic turbulence in strongly magnetized plasmas \cite{Brizard_Hahm_2007,Garbet_2010,Krommes_2012}. Its domain of application ranges from magnetic fusion experiments to the solar wind and the interstellar medium. As such, it is a theory capable of describing much of the matter in the observable universe. Although it is a kinetic model, gyrokinetics does not track the detailed evolution of the one-particle distribution function. Instead, the theory is formulated in terms of a quasi-particle distribution function defined on a reduced phase space. The quasi-particles, known as gyrocenters, move on a time scale that is long compared with the cyclotron period, and carry nontrivial electric and magnetic multipole moments \cite{Brizard_2008,Brizard_2009,Brizard_2013}.

Up to now \cite{Brizard_Hahm_2007}, gyrokinetic models have been derived by gyrocenter phase-space transformations that involved the perturbed electromagnetic potentials $(\Phi_{1}, {\bf A}_{1})$, in terms of which the perturbed electromagnetic fields ${\bf E}_{1} \equiv -\,\nabla\Phi_{1} - c^{-1}\partial{\bf A}_{1}/\partial t$ and ${\bf B}_{1} \equiv \nabla\btimes{\bf A}_{1}$ are defined. Since these potentials are not invariant under the gauge transformation $\Phi_{1} \rightarrow \Phi_{1} - c^{-1}\partial\chi_{1}/\partial t$ and ${\bf A}_{1} \rightarrow {\bf A}_{1} + \nabla\chi_{1}$, where $\chi_{1}({\bf x},t)$ represents an arbitrary scalar field, the current formulation of gyrokinetic theory (based on electromagnetic potentials) is not explicitly gauge independent (although it is gauge invariant). In particular, the equations of motion for individual gyrocenters involve the potentials in an essential manner.

The purpose of the present Letter is to derive a new gauge-free gyrokinetic theory, where the gyrocenter Hamiltonian is expressed explicitly in terms of simple integrals of the perturbed electromagnetic fields $({\bf E}_{1}, {\bf B}_{1})$. These gauge-free gyrokinetic equations can thus be used, for example, in hybrid kinetic models (e.g., Refs.~\cite{GeFi,Chen_Parker_2009}) in which one particle species is described in terms of a gauge-free gyrokinetic description and all other particle species are modeled in terms of a fully-kinetic particle description. In addition, the new equations will make it possible to calculate the motion of individual gyrocenters based on knowledge of the directly-observable electromagnetic fields. This capability should aid in the design and interpretation of new \emph{in situ} diagnostics for turbulent space and laboratory plasmas.

We begin our work by noting that the perturbed minimal-coupling electromagnetic Lagrangian $L_{1} = (e/c)\,{\bf A}_{1}\bdot\dot{\bf x} - e\,\Phi_{1}$ transforms as $L_{1} \rightarrow L_{1} + (e/c)\,d\chi_{1}/dt$ under a gauge transformation (in what follows, we consider particles of mass $m$ and charge $e$). Since Lagrangian mechanics is invariant under the addition of an exact time derivative to the Lagrangian, the Lagrangian $L_{1}$ is thus gauge-\emph{invariant}, although not gauge-\emph{free}. We will next consider how this basic gauge invariance property manifests itself after moving to the guiding center phase space, which is an intermediate step in the passage to gyrocenter phase space. 

Recall that the guiding center coordinate transformation \cite{Cary_Brizard_2009} is given by $(\mathbf{x},\mathbf{v})\mapsto (\mathbf{X},p_\parallel,J,\theta)\equiv \vb{z}_{\text{gc}}$, where $J=\mu B_0/\Omega_0$ is the guiding center gyroaction, and the guiding center position $\mathbf{X}$ is displaced from the particle position $\mathbf{x}$ by the guiding center gyroradius vector $\vb{\rho}(J,\theta)$ according to $\mathbf{x}({\bf X}, J,\theta) =\mathbf{X}+\vb{\rho}$. Instead of directly calculating $L_1$ in guiding center phase space, we perform a partial guiding-center transformation by replacing $\vb{\rho}$ with $\epsilon \vb{\rho}$, where $\epsilon$ is a dimensionless parameter ranging from $0$ (no shift in position) to $1$ (the full guiding center transformation). The partially-transformed minimal coupling Lagrangian is then given by 
\begin{equation}
 L_{1\epsilon} \;=\; (e/c)\,{\bf A}_{1\epsilon}\bdot(\dot{\bf X} + \epsilon\;\dot{\vb{\rho}}) - e\,\Phi_{1\epsilon} ,
\label{eq:gc_A1}
\end{equation}
where $\Phi_{1\epsilon} \equiv \Phi_{1}({\bf X} + \epsilon\vb{\rho},t)$ and ${\bf A}_{1\epsilon} \equiv {\bf A}_{1}({\bf X} + \epsilon\vb{\rho},t)$. Next, we observe that the parametric derivative $dL_{1\epsilon}/d\epsilon$ is given by
\begin{eqnarray*}
\frac{dL_{1\epsilon}}{d\epsilon} & = & e\,\vb{\rho}\bdot\left[{\bf E}_{1\epsilon} + \frac{1}{c}(\dot{\bf X} + \epsilon\,\dot{\vb{\rho}})\btimes{\bf B}_{1\epsilon}\right] \\
 &  &+\; \frac{d}{dt}\left(\frac{e}{c}\,\vb{\rho}\bdot{\bf A}_{1\epsilon}\right),\label{eq:para_deriv}
 \end{eqnarray*}
where ${\bf E}_{1\epsilon} \equiv -(\nabla\Phi_{1} + c^{-1}\partial{\bf A}_{1}/\partial t)_\epsilon$ and ${\bf B}_{1\epsilon} \equiv (\nabla\btimes{\bf A}_{1})_\epsilon$. We now use the simple identity $L_{1{\rm gc}} \equiv L_{1} + \int_{0}^{1} (dL_{1\epsilon}/d\epsilon)\,d\epsilon$, where the zero-Larmor-radius (ZLR) gauge-invariant Lagrangian $L_{1}  \equiv (e/c)\,{\bf A}_{1}({\bf X},t)\bdot\dot{\bf X} - e\,\Phi_{1}({\bf X},t)$ is evaluated at the guiding-center position ${\bf X}$, to obtain the alternative expression for $L_{1{\rm gc}}$:
\begin{eqnarray}
L_{1{\rm gc}}  & = & \frac{e}{c}\,\left( {\bf A}_{1} -\frac{}{} \vb{\rho}\btimes\mathbb{B}^{(0)}_{1}\right)\bdot\dot{\bf X} \;-\;  J\,\frac{\mathbb{B}_{1\parallel}^{(1)}}{B_0}\;\dot{\theta} \nonumber \\
 &  &-\; e\left( \Phi_{1} -\frac{}{} \vb{\rho}\bdot\mathbb{E}^{(0)}_{1}\right),
\label{eq:perturbed_gc}
\end{eqnarray}
where we have neglected the total time derivative and defined the multipole-integrated fields
\begin{equation}
\left(\begin{array}{c}
\mathbb{E}^{(n)}_{1}({\bf X},J,\theta) \\
\mathbb{B}^{(n)}_{1}({\bf X},J,\theta)
\end{array} \right) \equiv(n+1) \int_{0}^{1} \left(\begin{array}{c}
\epsilon^n{\bf E}_{1\epsilon} \\
\epsilon^n{\bf B}_{1\epsilon}
\end{array} \right) d\epsilon,
\label{eq:mathbb_EB}
\end{equation}
with $\mathbb{B}_{1\parallel}^{(1)} \equiv \bhat_{0}\bdot\mathbb{B}_{1}^{(1)}$. In the ZLR limit, we find $(\mathbb{E}^{(0)}_{1},\mathbb{B}^{(0)}_{1}) \simeq ({\bf E}_{1}, {\bf B}_{1})$ and $\mathbb{B}^{(1)}_{1} \simeq \mathbf{B}_1$. For the sake of simplicity in deriving Eq.~\eqref{eq:perturbed_gc}, we have used the assumption that the background magnetic field is uniform. There is no conceptual difficulty in extending this discussion to non-uniform background fields, although the calculations do become heavier.

Using the definitions \eqref{eq:perturbed_gc}-\eqref{eq:mathbb_EB}, the total perturbed guiding-center Lagrangian $L_{\rm gc} = L_{0{\rm gc}} + \epsilon_{\delta}\,L_{1{\rm gc}}$ is expressed in terms of the unperturbed guiding-center Lagrangian $L_{0{\rm gc}} =  (e/c)\,{\bf A}_{0}^{*}\bdot\dot{\bf X} + J\,\dot{\theta} - H_{0{\rm gc}}$, where the effective vector potential is given by $(e/c)\,{\bf A}_{0}^{*} \equiv (e/c)\,{\bf A}_{0} + p_{\|}\,\bhat_{0}$ (we ignore the higher-order gyrogauge terms here) and the guiding-center Hamiltonian is $H_{0{\rm gc}} = p_{\|}^{2}/2m + J\,\Omega_{0}$ (we ignore terms proportional to derivatives of the background magnetic field). We note that, while the unperturbed guiding-center Lagrangian $L_{0{\rm gc}}$ is independent of the gyroangle $\theta$, the perturbed part of the guiding-center Lagrangian \eqref{eq:perturbed_gc} is now written as the sum of the gyroangle-independent, gauge-invariant term $L_1$, and a manifestly gauge-free correction that is gyroangle dependent.

We will now use a gyrocenter phase-space transformation to remove the dependence on the gyroangle $\theta$ introduced by the gauge-free part of Eq.~\eqref{eq:perturbed_gc}. The ZLR minimal coupling Lagrangian introduces no $\theta$-dependence, so we may easily choose our transformation so that it does not alter this term. The strategy suggested here, i.e., focusing on the gauge-free part of the perturbed Lagrangian and leaving the minimal coupling term alone, proves to be the key to achieving a manifestly gauge-free theory to all orders in perturbation theory. 

 The result of the gyrocenter transformation is a new set of phase-space coordinates $(\ov{\bf X}, \ov{p}_{\|},\ov{J},\ov{\theta})\equiv \vb{z}_{\text{gy}}$, in which the gyrocenter Lagrangian is given by
\begin{equation}
L_{\rm gy} = \frac{e}{c}\,{\bf A}_{\rm gy}\bdot\dot{\ov{\bf X}} + \ov{J}\,\dot{\ov{\theta}} - H_{\rm gy},
 \label{eq:L_gy}
\end{equation}
where ${\bf A}_{\rm gy} \equiv \ov{\bf A}_{0}^{*} + \epsilon_{\delta}\,\ov{\bf A}_{1}$, and the gyroangle-independent gyrocenter Hamiltonian is given by
\begin{eqnarray}
H_{\rm gy} & = & K_{\rm gy} \;+\; \epsilon_{\delta}\,e\,\ov{\Phi}_{1}.
\label{eq:H_gy_def}
\end{eqnarray}
Here, the gyrocenter kinetic energy has the asymptotic expansion
\begin{equation}
K_{\text{gy}}=K_{0\text{gy}}+\epsilon_\delta\,K_{1\text{gy}}+\epsilon_\delta^2\,K_{2\text{gy}}+O(\epsilon_\delta^3),
\label{eq:K_gy}
\end{equation}
where $K_{0\text{gy}} = \ov{p}_{\|}^{2}/2m + \ov{\mu}\,\ov{B}_{0}$ denotes the lowest-order (guiding-center) kinetic energy. We will soon show explicitly that the higher-order terms $K_{n\text{gy}}$ $(n \geq 1)$ are gauge-free, which ensures that this new formulation of gyrokinetic theory is manifestly gauge-free. 

To first-order in $\epsilon_\delta$, the gauge-free gyrocenter transformation is given explicitly by $\vb{z}_{\text{gy}}=\vb{z}_{\text{gc}}-\epsilon_\delta\,\bm{\xi}_{1}$, where the first-order components are
\begin{eqnarray}
\xi_1^{\mathbf{X}} &=& \frac{\bhat_{0}}{m\Omega_{0}}\btimes\left(\nabla S_{1} - \mathbb{P}^{(0)}_{1}\right) \;+\; \bhat_{0}\;\pd{S_{1}}{p_{\|}}, \label{eq:xi_X} \\
\xi_1^{p_\parallel}  &=&  -\;\bhat_{0}\bdot\left(\nabla S_{1} - \mathbb{P}^{(0)}_{1}\right),  \label{eq:xi_p} \\
\xi_1^{J}  &=&  -\;\pd{S_{1}}{\theta} \;+\: J\left\langle\frac{\mathbb{B}_{1\parallel}^{(1)}}{B_0}\right\rangle,  \label{eq:xi_J} \\
\xi_1^{\theta}  &=& \pd{S_{1}}{J}.  \label{eq:xi_theta}
\end{eqnarray}
%\begin{eqnarray}
%\xi_{1}^{\bf X} & = & \frac{\bhat_{0}}{m\Omega_{0}}\btimes\left(\nabla S_{1} - \mathbb{P}^{(0)}_{1}\right) \;+\; \bhat_{0}\;\pd{S_{1}}{p_{\|}}, \label{eq:xi_X} \\
%\xi_{1}^{p_{\|}} & = & -\;\bhat_{0}\bdot\left(\nabla S_{1} - \mathbb{P}^{(0)}_{1}\right), \label{eq:xi_p} \\
%\xi_{1}^{J} & = & -\;\pd{S_{1}}{\theta} \;+\:  J\bhat_0\cdot\left\langle \frac{\mathbb{B}_1^{(1)}}{B_0}\right\rangle, \label{eq:xi_J} \\
%\xi_{1}^{\theta} & = & \pd{S_{1}}{J}, \label{eq:xi_theta}
%\end{eqnarray}
Here, $\mathbb{P}^{(0)}_{1} \equiv (e/c)\vb{\rho}\btimes\mathbb{B}^{(0)}_{1}$, and the gyroangle-dependent function $S_{1}$ is determined by the inhomogeneous linear equation
\begin{equation}
\frac{d_{0}S_{1}}{dt} = \left( \mu\,\mathbb{B}_{1\|}^{(1)}\right)^{\rm osc} - e\left[\vb{\rho}\bdot\left(\mathbb{E}^{(0)}_{1} + \frac{p_{\|}\bhat_{0}}{mc}\btimes\mathbb{B}^{(0)}_{1}\right)\right]^{\rm osc},
\end{equation}
where $Q^{\rm osc} \equiv Q - \langle Q\rangle$ denotes the gyroangle-dependent part of any function $Q$, and $d_{0}/dt \equiv \partial/\partial t + (p_{\|}/m)\,\bhat_{0}\bdot\nabla + \Omega_{0}\,\partial/\partial\theta$ is the uniform-background guiding-center evolution operator. 
%Here, we note that the function $S_{1}$ is determined by the gyroangle-dependent parts of the perturbed kinetic energy and the perturbed electric potential energy, respectively. We also note that, except for the exclusively gyroangle-dependent displacement \eqref{eq:xi_theta}, the displacements \eqref{eq:xi_X}-\eqref{eq:xi_J} have gyroangle-dependent parts (expressed in terms of $S_{1}$) and gyroangle-independent parts, which involve either $\langle \mathbb{P}^{(0)}_{1}\rangle$ or $\langle \mathbb{J}_{1\|}^{(1)}\rangle$ (both involving the perturbed magnetic field only). We can easily see, for example, that $(\xi_{1}^{p_{\|}})^{\rm osc} = -\,\bhat_{0}\bdot\nabla S_{1} \equiv m\,d_{0}\xi_{1\|}^{\bf X}/dt$, where $\xi_{1\|}^{\bf X} \equiv \bhat_{0}\bdot(\xi_{1}^{\bf X})^{\rm osc}$ is exclusively 
%gyroangle-dependent.

The first two terms in the asymptotic expansion \eqref{eq:K_gy} of the gyrocenter kinetic energy are given by
\begin{equation}
K_{1\text{gy}} = \ov{\mu}\,\left\langle\ov{\mathbb{B}}^{(1)}_{1\|}\right\rangle-\left\langle e\overline{\bm{\rho}}\bdot\left(\ov{\mathbb{E}}^{(0)}_{1} + \frac{\mathbf{v}_{0\text{gy}}}{c}\times\ov{\mathbb{B}}^{(0)}_{1}\right)\right\rangle,
\label{eq:first_order} 
%K_{2\text{gy}} &= & -\,\left\langle\ov{\xi_{1}}^{\bf X}\right\rangle\bdot\left[ e\,\left(\ov{\bf E}_{1} + \frac{\ov{p}_{\|}\bhat_{0}}{mc}\btimes\ov{\bf B}_{1} \right) - \ov{\nabla}K_{1{\rm gy}}\right] \nonumber \\
% &  &+\; \frac{1}{2m} \left( \left\langle\left(\bhat_{0}\bdot\ov{\mathbb{P}}^{(0)}_{1}\right)^{2}\right\rangle \;+\; \bhat_{0}\bdot\left\langle\ov{\vb{\Pi}}_{1}\btimes\pd{\ov{\vb{\Pi}}_{1}}{\ov{\theta}}\right\rangle\right) \nonumber \\
%  & &+\; \left\langle\ov{\mathbb{J}}_{1\|}^{(1)}\right\rangle\; \pd{K_{1{\rm gy}}}{\ov{J}} \;+\; \left\langle \ov{\xi_{1}}^{J}\;\pd{\ov{\xi_{1}}^{\theta}}{\ov{\theta}}\right\rangle\;\Omega_{0},
%\label{eq:second_order}
\end{equation}
where $\mathbf{v}_{\text{gy}0}= (\ov{p}_{\|}/m)\,\bhat_0$ is the leading-order gyrocenter velocity [see Eq.~\eqref{eq:X_dot}], and
\begin{eqnarray}
K_{2\text{gy}} & = & \ov{J}\;\frac{\Omega_{1{\rm gy}}}{B_0}\left\langle\ov{\mathbb{B}}_{1\parallel}^{(1)}\right\rangle + \left\langle\ov{\mathbb{P}}_{1}^{(0)}\right\rangle\bdot\left( {\bf v}_{{\rm gy}1} - 
\frac{\bhat_{0}}{m}\;\left\langle\ov{\mathbb{P}}_{1\|}^{(0)}\right\rangle \right) \nonumber \\
 &  &+\; \frac{1}{2m}\;\left\langle \left(\ov{\mathbb{P}}_{1\|}^{(0)}\right)^{2}\right\rangle + \frac{m}{2}\,\bhat_{0}\bdot\left\langle \Omega_{0}\xi_{1}^{\bf X}\btimes\frac{d_{0}\xi_{1}^{\bf X}}{dt}\right\rangle \nonumber \\
  &  &+\; \frac{e}{\Omega_{0}}\left\langle \vb{\rho}\bdot\left(\ov{\mathbb{E}}^{(0)}_{1} + \frac{\mathbf{v}_{0\text{gy}}}{c}\times\ov{\mathbb{B}}^{(0)}_{1}\right)\;\frac{d_{0}\xi_{1}^{\theta}}{dt}\right\rangle,
\label{eq:second_order}
\end{eqnarray}
where $\Omega_{1{\rm gy}} \equiv \partial K_{1\text{gy}}/\partial\ov{J} = -\;\langle \ov{\bf F}_{1{\rm gc}}\bdot\partial\ov{\vb{\rho}}/\partial\ov{J}\rangle$ and
\[ \mathbf{v}_{\text{gy}1}= \left\langle\ov{\mathbb{P}}_{1\|}^{(0)}\right\rangle\frac{\bhat_0}{m}+ \left( \langle\ov{\bf F}_{1\text{gc}}\rangle \;-\; \frac{d_{0}\langle\ov{\mathbb{P}}^{(0)}_{1}\rangle}{dt}\right)\times\frac{\bhat_0}{m\Omega_0} \]
are the first-order corrections to the gyrofrequency and gyrocenter velocity, respectively, with
\[ {\bf F}_{1\text{gc}} =e\,\mathbf{E}_{1\text{gc}}+\frac{e}{c}\left(\frac{p_{\|}}{m}\bhat_0+\Omega_{0}\pd{\vb{\rho}}{\theta}\right)\times\mathbf{B}_{1\text{gc}}. \]
We note that, in the standard (low-frequency) gyrokinetic ordering, we can approximate $d_{0}/dt \simeq \Omega_{0}\,\partial/\partial\ov{\theta}$ in Eq.~\eqref{eq:second_order}. In the ZLR limit (for a uniform magnetic field), we obtain the simplified expression for Eq.~\eqref{eq:second_order}:
\begin{eqnarray}
K_{2\text{gy}} & \simeq & -\; \frac{mc^{2}}{2\,B_{0}^{2}}\,|\ov{\bf E}_{1\bot}|^{2}  \;+\; \ov{p}_{\|}\,\frac{c\bhat_{0}}{B_{0}^{2}}\bdot\ov{\bf E}_{1}\btimes\ov{\bf B}_{1} \nonumber \\
  &  &+\; \left( \ov{\mu}\,B_{0} - \frac{\ov{p}_{\|}^{2}}{m}\right)\,\frac{|\ov{\bf B}_{1\bot}|^{2}}{2\,B_{0}^{2}},
\label{eq:K2gy_simple}
\end{eqnarray}
which is a standard result in gyrokinetic theory \cite{Brizard_Hahm_2007}.
%is the first-order correction to the gyrofrequency, 
%\begin{align}
%\mathbf{v}_{\text{gy}1}=\ov{\tilde{v}}_{\parallel 1}\bhat_0+\frac{\langle\ov{\mathbf{F}}_{\text{gc}}\rangle\times\bhat_0}{m\Omega_0}
%\end{align}
%is the first-order correction to the gyrocenter velocity, and 
%\begin{align}
%\mathbf{U}_{0\text{gy}}=\Omega_0\frac{\partial}{\partial\ov{\theta}}+\ov{v}_\parallel\bhat_0\cdot\frac{\partial}{\partial\ov{\mathbf{X}}}
%\end{align}
%is the leading-order Eulerian gyrocenter phase space fluid velocity. 
%where the first-order perturbed momentum 
%\[ \ov{\vb{\Pi}}_{1} = \ov{\nabla}S_{1} - \ov{\mathbb{P}}^{(0)}_{1} \equiv \ov{\xi_{1}}^{p_{\|}}\,\bhat_{0} + m\Omega_{0}\,\ov{\xi_{1}}^{\bf X}\btimes\bhat_{0} \]
%is defined in terms of the first-order displacements \eqref{eq:xi_X}-\eqref{eq:xi_p}, and we have approximated $d_{0}/dt \simeq \Omega_{0}\,\partial/\partial\ov{\theta}$ in Eq.~\eqref{eq:second_order}. 

The physical meaning of the first-order terms in Eq.~\eqref{eq:first_order} is given in terms of lowest-order gyrocenter electric and magnetic multipole contributions: $\ov{\mu}\,\langle\ov{\mathbb{B}}^{(1)}_{1\|}\rangle$ yields the intrinsic multipole moment contribution to magnetization while $-\,\langle e\ov{\vb{\rho}}\bdot[\ov{\mathbb{E}}^{(0)}_{1} + (\ov{p}_{\|}\bhat_{0}/mc)\btimes\ov{\mathbb{B}}^{(0)}_{1}]\rangle$ yield the electric multipole moment contribution to polarization and the moving electric-multipole moment contribution to magnetization, respectively [see Eq.~\eqref{eq:polmag_0}]. The physical meaning of the second-order terms in Eq.~\eqref{eq:second_order} is given in terms of higher-order contributions to polarization and intrinsic magnetization [e.g., first and third terms in Eq.~\eqref{eq:K2gy_simple}] and the moving-electric-dipole contribution to magnetization [e.g., second term in Eq.~\eqref{eq:K2gy_simple}].

By applying Hamilton's principle to the gyrocenter Lagrangian \eqref{eq:L_gy}, phase-space-conserving gyrocenter equations of motion may be derived. In their general form, these consist of the gyrocenter velocity equation
\begin{equation}
\mathbf{v}_{\text{gy}} = \pd{K_{\rm gy}}{\ov{p}_{\|}}\frac{{\bf B}_{\rm gy}}{B_{\|{\rm gy}}} + \left( \epsilon_{\delta}\;e\,\ov{\bf E}_{1} - \ov{\nabla} K_{\rm gy}\right)\btimes\frac{c\bhat_{0}}{eB_{\|{\rm gy}}},
\label{eq:X_dot}
\end{equation}
and the gyrocenter parallel-force equation
\begin{equation}
m\,a_{\parallel\text{gy}} = \frac{{\bf B}_{\rm gy}}{B_{\|{\rm gy}}}\bdot\left( \epsilon_{\delta}\,e\,\ov{\bf E}_{1} \;-\frac{}{} \ov{\nabla} K_{\rm gy}\right),
\label{eq:p_dot}
\end{equation}
where $B_{\|{\rm gy}} \equiv \bhat_{0}\bdot{\bf B}_{\rm gy}\equiv \bhat_0\bdot(\ov{\mathbf{B}}_{0}^{*}+\epsilon_\delta\ov{\mathbf{B}}_1)$ appears in the definition of the gyrocenter Jacobian ${\cal J}_{\rm gy} \equiv (e/c)\,B_{\|{\rm gy}}$. Note that, by construction, the gyrocenter gyroaction $\ov{J}$ is conserved by the Hamiltonian gyrocenter dynamics (although it is still an adiabatic invariant with respect to the exact Hamiltonian particle dynamics), and the gyrocenter evolution for the gyroangle $\ov{\theta}$ is decoupled from the reduced gyrocenter dynamics represented by Eqs.~\eqref{eq:X_dot}-\eqref{eq:p_dot}. In addition, the gyrocenter Jacobian ${\cal J}_{\rm gy}$ satisfies the gyrocenter Liouville Theorem: $\partial{\cal J}_{\rm gy}/\partial t + \ov{\nabla}\bdot(\mathbf{v}_{\text{gy}}\,{\cal J}_{\rm gy}) + \partial(m\,a_{\parallel\text{gy}}\,{\cal J}_{\rm gy})/
\partial\ov{p}_{\|} = 0$.

As is well-known in the context of equilibrium thermodynamics, the polarization or magnetization of a material is given by differentiating the material's free energy with respect to the electric or magnetic field
\cite{Landau_Lifshitz}. In Refs.\,\cite{Morrison_2013,Burby_2015}, the non-equilibrium analogue of this fact is shown to involve \emph{functional} derivatives of the net gyrocenter kinetic energy $\mathcal{K}_{\text{gy}}= \int_{\text{gy}} K_{\text{gy}}\,\ov{F}$, which may therefore be regarded as a non-equilibrium analogue of the free energy of a gyrocenter gas. Here, $\int_{\text{gy}} \equiv \sum \int d\bm{z}_{\text{gy}}$ denotes a sum over particle species and an integration over gyrocenter phase space, and the gyrocenter distribution $\ov{F}$ includes the gyrocenter Jacobian ${\cal J}_{\rm gy}$. In particular, the polarization and magnetization densities $(\bm{\mathcal{P}}_{\text{gy}},\bm{\mathcal{M}}_{\text{gy}})$ of a distribution of gyrocenters $\ov{F}$ are given by 
\begin{equation}
\left(\begin{array}{c}
\bm{\mathcal{P}}_{\text{gy}}({\bf x}) \\
\bm{\mathcal{M}}_{\text{gy}}({\bf x}) 
\end{array} \right) = \int_{\text{gy}} \left( \begin{array}{c}
\vb{\pi}_{\rm gy}({\bf x}) \\
\vb{\mu}_{\rm gy}({\bf x})
\end{array} \right) \ov{F},
\label{eq:PM_gy}
\end{equation}
where ${\bf x}$ denotes an arbitrary field point, while $\vb{\pi}_{\rm gy}({\bf x}) \equiv -\,\epsilon_{\delta}^{-1}\delta K_{\text{gy}}/\delta\mathbf{E}_{1}({\bf x})$ and $\vb{\mu}_{\rm gy}({\bf x}) \equiv -\,
\epsilon_{\delta}^{-1}\delta K_{\text{gy}}/\delta\mathbf{B}_{1}({\bf x})$ are the gyrocenter electric and magnetic dipole moments, respectively, derived from the gyrocenter Hamiltonian \eqref{eq:K_gy}. Using the first-order Hamiltonian \eqref{eq:first_order}, we find 
\begin{equation}
\left. \begin{array}{l}
\vb{\pi}_{0{\rm gy}}({\bf x}) = e\,\langle\ov{\vb{\rho}}\,\delta_{\bf x}^{(0)}\rangle \\
\vb{\mu}_{0{\rm gy}}({\bf x}) = -\,\ov{\mu}\,\bhat_{0}\,\langle\delta_{\bf x}^{(1)}\rangle + e\,\langle\ov{\vb{\rho}}\,\delta_{\bf x}^{(0)}\rangle\btimes{\bf v}_{0{\rm gy}}/c 
\end{array} \right\},
\label{eq:polmag_0}
\end{equation}
where $\delta_{\bf x}^{(n)} = (n+1)\int_{0}^{1}\epsilon^{n}\delta^{3}(\ov{\bf X} + \epsilon\ov{\vb{\rho}} - {\bf x})\,d\epsilon$. Using the second-order Hamiltonian \eqref{eq:second_order}, on the other hand, we find the first-order corrections
\begin{eqnarray}
\vb{\pi}_{1{\rm gy}} & = & \left\langle\pounds_{\vb{\xi}}\left(e\,\ov{\vb{\rho}}\,\delta_{\bf x}^{(0)}\right)\right\rangle, 
\label{eq:pi_1} \\
\vb{\mu}_{1{\rm gy}} & = & -\,\ov{\mu}\,\bhat_{0}\,\langle\delta_{\bf x}^{(1)}\rangle\;\frac{\Omega_{1{\rm gy}}}{\Omega_{0}} + \frac{e}{c}\,\left\langle \ov{\vb{\rho}}\,\delta_{\bf x}^{(0)}\right\rangle\btimes{\bf v}_{1{\rm gy}}  
\label{eq:mu_1} \\
 &  &+ \left\langle \pounds_{\vb{\xi}}\left( -\,\ov{\mu}\,\bhat_{0}\,\delta_{\bf x}^{(1)} + \frac{e}{c}\,\ov{\vb{\rho}}\,\delta_{\bf x}^{(0)}\btimes{\bf v}_{0{\rm gy}} \right) \right\rangle, \nonumber
\end{eqnarray}
where the first-order Lie derivative $\pounds_{\vb{\xi}} \equiv \xi_{1}^{\bf X}\bdot\ov{\nabla} + \xi_{1}^{p_{\|}}\,\partial_{\ov{p}_{\|}} + \xi_{1}^{J}\,\partial_{\ov{J}} + \xi_{1}^{\theta}\,\partial_{\ov{\theta}}$ is defined in terms of the first-order components \eqref{eq:xi_X}-\eqref{eq:xi_theta} and contributions from $B_{1\|}$ are omitted in Eqs.~\eqref{eq:pi_1}-\eqref{eq:mu_1} since they appear at higher-order in the FLR expansion [i.e., $B_{1\|}$ is absent in Eq.~\eqref{eq:K2gy_simple}]. Here, at each order, the magnetic dipole is given as the sum of the intrinsic part (first term) and the moving electric dipole part (second term). 

The polarization and magnetization densities, together with the gyrocenter equations of motion, completely specify our new gauge-free gyrokinetic theory. The gyrocenter distribution function $\ov{F}$ obeys the gyrokinetic equation
\begin{align}
\partial\ov{F}/\partial t + \ov{\nabla}\cdot(\mathbf{v}_{\text{gy}}\,\ov{F})+\partial(m \,a_{\parallel\text{gy}} \ov{F})/\partial\ov{p}_{\|}=0.\label{eq:gk_equation}
\end{align}
The electromagnetic fields obey Maxwell's equations with charge and current densities $\varrho({\bf x}) = \varrho_{\text{gy}}({\bf x}) - \nabla\cdot\bm{\mathcal{P}}_{\text{gy}}({\bf x})$ and $\mathbf{J}({\bf x}) =  
{\bf J}_{\rm gy}({\bf x}) + c\;\nabla\times\bm{\mathcal{M}}_{\text{gy}}({\bf x}) + \partial_t\bm{\mathcal{P}}_{\text{gy}}({\bf x})$, where the gyrocenter contributions 
\begin{equation}
\left. \begin{array}{rcl}
\varrho_{\text{gy}}({\bf x}) & = & \int_{\text{gy}} \ov{F}\;e\,\delta(\ov{\bf X} - {\bf x}) \\
{\bf J}_{\rm gy}({\bf x}) & = & \int_{\text{gy}}  \ov{F}\;e\,\mathbf{v}_{\text{gy}}\,\delta(\ov{\bf X} - {\bf x})
\end{array} \right\}
\end{equation}
include the delta function $\delta(\ov{\bf X} - {\bf x}) \equiv \delta_{0}$, thereby guaranteeing that only gyrocenters located at the field position $\ov{\bf X} = {\bf x}$ contribute to the charge and current densities. We note that the gyrocenter velocity $\mathbf{v}_{\text{gy}}$, given by Eq.~\eqref{eq:X_dot}, includes first-order and second-order contributions through the gyrocenter kinetic energy \eqref{eq:K_gy}.

By using the identity $\ov{\vb{\rho}}\bdot\ov{\nabla}\delta^{(0)}_{\bf x} \equiv \delta_{\rm gc} - \delta_{0}$, where $\delta_{\rm gc} \equiv \delta(\ov{\bf X} + \ov{\vb{\rho}} - {\bf x})$, the charge density becomes
\begin{equation}
\varrho({\bf x}) \;=\; \int_{\rm gy} \ov{F} \left[ e\,\langle\delta_{\rm gc}\rangle \;+\; \epsilon_{\delta}\;\left\langle\pounds_{\vb{\xi}}\left(e\;\delta_{\rm gc}\right)\right\rangle \right],
\label{eq:rho_final}
\end{equation}
which corresponds exactly to the standard gyrokinetic expression \cite{Brizard_Hahm_2007}, except that the generating vector-field components \eqref{eq:xi_X}-\eqref{eq:xi_theta} in Eq.~\eqref{eq:rho_final} are now gauge-free. Next, we use the identity $\ov{\mu}\,\bhat_{0}\btimes\ov{\nabla}\delta_{\bf x}^{(1)} \equiv -\,e\,\delta_{\rm gc}\,\ov{\bf v}_{\bot}/c + \Omega_{0}\partial_{\ov{\theta}}(e\,\delta_{\bf x}^{(0)}\ov{\vb{\rho}}/c)$, and we find the current density
\begin{eqnarray}
{\bf J}({\bf x}) & = & \int_{\rm gy} \ov{F} \left[ \frac{}{} e \left({\bf v}_{0{\rm gy}} + \epsilon_{\delta}\,{\bf v}_{1{\rm gy}}\right) \langle \delta_{\rm gc}\rangle + \epsilon_{\delta}^{2}\;e\,{\bf v}_{2{\rm gy}}\;\delta_{0} \right. \nonumber \\
 & &+\; e\,\left(\Omega_{0} \,+\frac{}{} \epsilon_{\delta}\,\Omega_{1{\rm gy}}\right) \left\langle\pd{\ov{\vb{\rho}}}{\ov{\theta}}\; \delta_{\rm gc}\right\rangle 
 \label{eq:J_final} \\
 & &\left.+  e\,\epsilon_{\delta} \left\langle \pounds_{\vb{\xi}}\left[ \left( {\bf v}_{0{\rm gy}} + \ov{\bf v}_{\bot} \right) \delta_{\rm gc} - \Omega_{0}\partial_{\ov{\theta}}(\ov{\vb{\rho}}\delta_{\bf x}^{(0)}) \right] \right\rangle \right], \nonumber
\end{eqnarray}
where the gyrocenter polarization current density $\partial_{t}\bm{\mathcal{P}}_{\rm gy}$ has been omitted as a higher-order correction. Approximate expressions for the charge and current densities \eqref{eq:rho_final}-\eqref{eq:J_final} can be used in the truncated delta-$f$ formulation of gyrokinetic theory:
\begin{align}
\varrho
\approx& \int_{\rm gy}\left\langle e\,\delta_{\text{gc}}\right\rangle\,\ov{F} +\epsilon_\delta \int_{\rm gy}\left\langle\left(\frac{e\,\overline{\vb{\rho}}\cdot\overline{\mathbb{E}}_1^{(0)}}{T}\right)^{\text{osc}}\,e\,\delta_{\text{gc}}\right\rangle\,\ov{F}_M \\
\mathbf{J}
\approx& \int_{\rm gy} \left\langle e\,({\bf v}_{0{\rm gy}}+\overline{\mathbf{v}}_\perp)\,\delta_{\text{gc}}\right\rangle\,\ov{F} \nonumber\\
&+\epsilon_\delta  \int_{\rm gy} e\,\left( \frac{\Omega_{1{\rm gy}}}{\Omega_0}\left\langle\overline{\mathbf{v}}_\perp\,\delta_{\text{gc}}\right\rangle + \mathbf{v}_{1\text{gy}}\,\langle\delta_{\text{gc}}\rangle\right)\,\ov{F}_M,
\end{align}
where the gyrocenter Vlasov distribution $\ov{F}$ is approximated by a local Maxwellian distribution $\ov{F}_M$ whenever first-order effects are considered, and second-order effects are omitted. Complete variational derivations of full and truncated gauge-free gyrokinetic models will be presented in a future publication. Here, we simply note that the total energy (Hamiltonian) functional for the gauge-free gyrokinetic Vlasov-Maxwell equations is expressed as \cite{Burby_2015}
\begin{eqnarray}
{\cal H}_{\rm gy} & = & \int_{\rm gy} \ov{F} \left( K_{\rm gy} \;-\; {\bf E}_{1}\bdot\fd{K_{\rm gy}}{{\bf E}_{1}}\right) \nonumber \\
 &  &+\; \int\frac{d^{3}x}{8\pi}\left(\epsilon_{\delta}^{2}\,|{\bf E}_{1}|^{2} \;+\frac{}{} |{\bf B}|^{2}\right),
\end{eqnarray}
where ${\bf B} = {\bf B}_{0} + \epsilon_{\delta}\,{\bf B}_{1}$ denotes the total magnetic field.

An important application of these results is building a gauge-free hybrid gyrokinetic electron-Lorentz ion model. As was explained for instance in Ref.\,\cite{Chen_Parker_2009}, present-day turbulence simulations that employ gyrokinetic electrons and ions are often forced to resolve timescales that are near the ion cyclotron period. Therefore, not much is gained, and much is lost, when applying gyrokinetics to both species. A hybrid model would have the obvious advantages of simpler equations for the ions and the capability of resolving the ion-cyclotron range of frequencies. While there is an established hybrid model described in Ref.\,\cite{GeFi}, it is based on the electromagnetic potentials instead of the fields. The success of the gauge-free hybrid drift-kinetic electron-Lorentz ion model formulated in Ref.\,\cite{Chen_Parker_2009} motivates extending the gauge-free approach beyond the drift-kinetic approximation. Such a model will be the subject of future publications.

In summary, the gauge-free formulation of gyrokinetic theory presented here has an underlying Hamiltonian structure that was described in Ref.\,\cite{Burby_2015}, as well as a variational structure in the spirit of Refs.\,\cite{Sugama_2000,Brizard_2000a,Brizard_2000b,Brizard_Tronci_2016}. Therefore the formalism introduced in Ref.\,\cite{Burby_2017} may be used to develop finite-dimensional truncations of our new theory that also possess a Hamiltonian structure. Such truncations may be used to develop structure-preserving schemes for simulating kinetic turbulence, much as was done for the Vlasov-Maxwell system in Refs.\,\cite{Xiao_2015, He_2016, Kraus_2017}. 

%We must emphasize, however, that the approximate expressions we have given for the charge and current densities omit some higher-order terms that would appear in a rigorously Hamiltonian version of our theory. We have opted to omit these terms here because they lead to more cumbersome expressions, and not because of any essential difficulty they introduce into the theory.

This research was supported by the U.~S.~Department~of~Energy, Office of Science, Fusion Energy Sciences under  Award No. DE-FG02-86ER53223  and the U.S. Department of Energy Fusion Energy Sciences Postdoctoral Research Program administered by the Oak Ridge Institute for Science and Education (ORISE) for the DOE. ORISE is managed by Oak Ridge Associated Universities (ORAU) under DOE contract number DE-AC05-06OR23100. All opinions expressed in this paper are the author's and do not necessarily reflect the policies and views of DOE, ORAU, or ORISE. The present work was also partially funded by grants (AJB) from the U.~S.~Department~of~Energy under contract DE-SC0014032 and the National Science Foundation, grant number PHY-1805164.

\end{document}